\documentclass[%
 reprint,
amsmath,amssymb,
aps,
]{revtex4-1}
\usepackage{graphicx}
\usepackage{dcolumn}
\usepackage{bm}
\usepackage{epstopdf}
\usepackage[dvipsnames]{xcolor}

\begin{document}

\preprint{APS/123-QED}

\title{Fully gapped superconductivity without sign reversal in the topological superconductor PbTaSe$_2$}

\author{Yue Sun,$^{1,2}$}
\email{sunyue@phys.aoyama.ac.jp}
\author{Shunichiro Kittaka,$^{1,8}$ Toshiro Sakakibara,$^1$ Kazushige Machida,$^3$ R. Sankar,$^4$ Xiaofeng Xu$^5$, Nan Zhou,$^{1,6}$ Xiangzhuo Xing$^6$, Zhixiang Shi$^6$, Sunseng Pyon$^7$ and Tsuyoshi Tamegai$^7$}

\affiliation{%
	$^1$Institute for Solid State Physics (ISSP), The University of Tokyo, Kashiwa, Chiba 277-8581, Japan\\
	$^2$Department of Physics and Mathematics, Aoyama Gakuin University, Sagamihara 252-5258, Japan\\
	$^3$Department of Physics, Ritsumeikan University, Kusatsu, Shiga 525-8577, Japan\\
	$^4$Institute of Physics, Academia Sinica, Nankang, Taipei 11529, Taiwan\\
	$^5$Department of Applied Physics, Zhejiang University of Technology, Hangzhou 310023, China\\
	$^6$School of Physics and Key Laboratory of MEMS of the Ministry of Education, Southeast University, Nanjing 211189, China\\
	$^7$Department of Applied Physics, The University of Tokyo, Bunkyo-ku, Tokyo 113-8656, Japan\\
    $^8$Department of Physics, Chuo University, Bunkyo-ku, Tokyo 112-8551, Japan}

\date{\today}

\begin{abstract}
We investigate the superconducting gap function of topological superconductor PbTaSe$_2$. Temperature, magnetic field, and three-dimensional (3D) field-angle dependences of the specific heat prove that the superconductivity of PbTaSe$_2$ is fully-gapped, with two isotropic $s$-wave gaps. The pair-breaking effect is probed by systematically increasing non-magnetic disorders through H$^+$-irradiations. The superconducting transition temperature, $T_{\rm{c}}$, is found to be robust against disorders, which suggests that the pairing should be sign-preserved rather than sign-reversed.       

\end{abstract}

\maketitle
Topological superconductivity (TSC) has stimulated great interests in condensed matter physics \cite{QiRevModPhys.83.1057,Sato_2017review}. It is partly due to the possibility of hosting Majorana bound state, which obeys the non-Abelian statistics, and can be used in topological quantum computation \cite{KITAEV20032AnnalsofPhysics,NayakRevModPhys.80.1083}. In real materials, the topological superconductors can be realized in two primary routes. One way is to make heterostructure between conventional $s$-wave superconductors and topological insulators \cite{LiangFuPhysRevLett.100.096407}, quantum anomalous Hall insulators \cite{QiPhysRevB.82.184516}, or ferromagnetic atomic chains \cite{Nadj-PergePhysRevB.88.020407}, where the proximity effect on a spin non-degenerate band induces TSC on the interface. Another way is to search materials naturally hosting TSC \cite{QiRevModPhys.83.1057,Beenakkerannurev}. In the second route, both the bulk TSC with spin-triplet $p$-wave pairing, and the surface TSC above a bulk spin-singlet $s$-wave superconductivity are realized \cite{Sato_2017review,QiRevModPhys.83.1057,LiangFuPhysRevLett.100.096407,SasakiPhysRevLett.107.217001,FuLiangPRL,ZhangNatPhy}. In the second route, the noncentrosymmetric superconductors are one of the most promising way \cite{SatoPhysRevB.79.094504,Yipannurev}. Due to the broken inversion symmetry, spin-orbital coupling becomes antisymmeric, which can lift spin degeneracy, and induce the topological state. It allows the mixing of spin-singlet and spin-triplet pairing, suggesting both the bulk and surface TSC are hopeful to be realized \cite{GuanPbTaSeSciAdv}.                 

Noncentrosymmetric superconductor PbTaSe$_2$ shows a bulk superconducting transition temperature, $T_{\rm{c}}$, $\sim$ 3.7 K, with a strong spin-orbital coupling, which results in large Rashba splitting and the breaking of spin degeneracy \cite{AliPRB}. Recent angle-resolved photoemission spectroscopy (ARPES) measurements reveals the existence of topological nodal lines near the Fermi surface, which are protected by the reflection symmetry \cite{BianPbTaSeNatComm}. The topological state is also confirmed by the following scanning tunneling microscopy (STM) measurements \cite{GuanPbTaSeSciAdv}, and a zero-energy bound state at the vortex core was also observed \cite{GuanPbTaSeSciAdv,ZhangSTMMajorana}. These results suggest that PbTaSe$_2$ is a promising candidate for TSC \cite{BianPbTaSeNatComm,GuanPbTaSeSciAdv,ZhangSTMMajorana,ChangPbTaSePRB,XuCQPbTaSePRB,XuXitongPbTaSePRB,LianPRB}. 

To reveal the origin and mechanism of the TSC in PbTaSe$_2$, it is crucial to probe its bulk gap function. Up to now, no-breaking of time-reversal symmetry has been reported based on the muon spin rotation and relaxation ($\mu$SR) experiments \cite{WilsonPbTaSePRB}. A fully-gapped structure has been suggested by the measurements of specific heat \cite{ZhangPbTaSePRB}, thermal conductivity \cite{WangPbTaSePRB}, penetration depth \cite{PangPRBpenetration}, and nuclear magnetic resonance (NMR) \cite{MaedPbTaSePRB}, while details about the gap structure such as the multi-gap or single-gap, isotropic or anisotropic are still under controversy \cite{WangPbTaSePRB,PangPRBpenetration,LianPRB}. Besides, most evidences for the fully-gapped structure are obtained from the excited quasi-particles (QPs) in the $ab$-plane since the applied magnetic fields are along the $c$-axis \cite{ZhangPbTaSePRB,WangPbTaSePRB,PangPRBpenetration}. Hence, the $p$-wave-like gap structure with nodes along the $k_z$-directions, which can host Majorana fermions in its side surface, has not been well identified \cite{KobayashiPRL,HashimotoPRB,YangPRL}. Furthermore, even for the fully-gapped structure, the pairing can be sign-reversed or sign-preserved. A well known example is the iron-based superconductors, where the gap function has no nodes but may still change sign between the electron and hole pockets \cite{Mazinnature,HirschfeldReportsonProgressinPhysics}. To probe the gap structure including possible sign reversal, a bulk technique capable of probing QPs excitations with 3D angle resolution, and a phase-sensitive probe are needed.

\begin{figure*}\center
	\includegraphics[width=16cm]{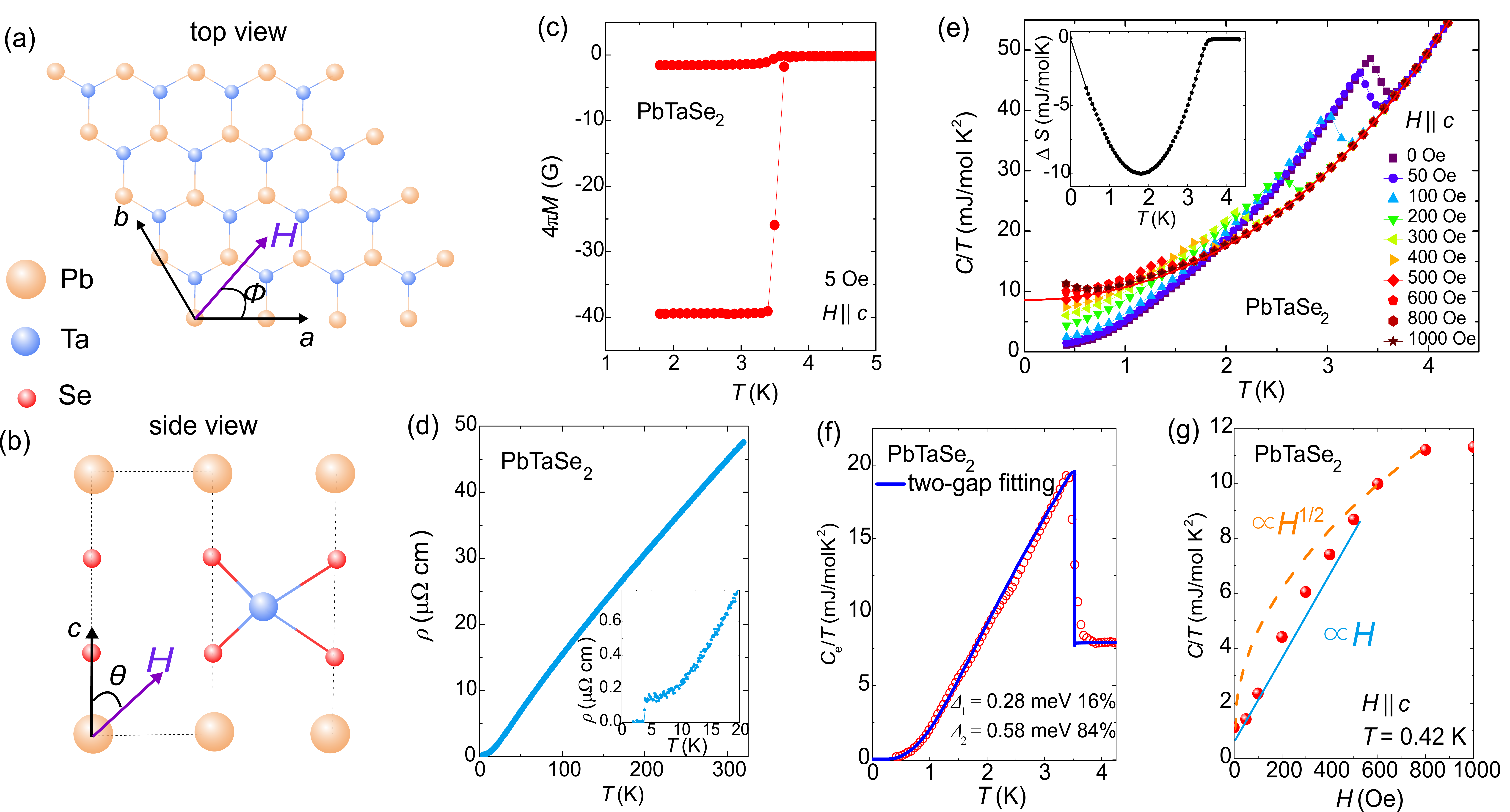}\\
	\caption{Crystal structure of PbTaSe$_2$ in the (a) hexagonal plane from top view, and (b) the noncentrosymmetric structure from the side view. The orange, blue, and red circles represent the elements of Pb, Ta, and Se, respectively. The definitions of azimuthal angle $\phi$ and polar angle $\theta$ with respect to the crystal structure are also shown. (c) Temperature dependence of the magnetization under 5 Oe field perpendicular to the hexagonal plane. (d) Temperature dependence of the in-plane resistivity measured at zero field. Inset is the enlarged plot below 20 K. (e) Temperature dependence of specific heat plotted as $C/T$ vs $T$ under magnetic fields ($H\parallel c$ ) ranging from 0 to 1000 Oe. The solid line represents the fit to the normal state specific heat. The inset shows the temperature dependence of the difference between the superconducting and normal states entropies, $\Delta S$. (f) Zero-field electronic specific heat $C_e/T$ vs $T$, together with the fit by a two isotropic $s$-wave gap model. (g) Magnetic field dependence of specific heat at 0.42 K for $H \parallel c$. The solid and dashed lines represent the $H$ and $H^{1/2}$ dependences.}\label{}
\end{figure*}

In this report, we combined the measurements of 3D field angle-resolved specific heat (ARSH) and the disorder effects induced by H$^+$-irradiation to probe the gap function of PbTaSe$_2$. The former technique has a 3D angle resolution of QPs, while the latter one is phase-sensitive. The superconductivity of PbTaSe$_2$ is found to be fully-gapped, consisting of two isotropic $s$-wave gaps with sign-preserved pairing.                      

PbTaSe$_2$ single crystals were grown by the chemical vapor transport method \cite{SankarPbTaSeJPCM}. Magnetization measurements were performed using a commercial SQUID magnetometer (MPMS-XL5, Quantum Design). The resistivity were measured by the four-probe method in a physical property measurement system (PPMS, Quantum Design). The temperature dependence of the specific heat under various magnetic fields was also measured by using PPMS. The 3D field-angle dependence of the specific heat was measured in an 8 T split-pair superconducting magnet with a $^3$He refrigerator. The refrigerator can be continuously rotated by a motor on top of the dewar with an angular resolution better than 0.01$^\circ$. The calibration and validity of the measurement system can be seen in our previous report \cite{SunPhysRevLett.123.027002}. Single crystals used for the irradiation experiments were cleaved to thin plates with thickness $\sim$30 $\mu$m along the $c$-axis, which is smaller than the projected range of 3-MeV H$^+$ for PbTaSe$_2$ of $\sim$47 $\mu$m \cite{irradiationrange}. To avoid possible sample dependent influence, all the measurements were performed on one identical piece of crystal, which was divided into several pieces and irradiated by H$^+$ up to doses of 0 (pristine), 0.5$\times$10$^{16}$, 1$\times$10$^{16}$, 2$\times$10$^{16}$, 4$\times$10$^{16}$, and 8$\times$10$^{16}$/cm$^2$, respectively. We estimated that 1$\times$10$^{16}$ dose H$^+$-irradiation is supposed to cause about 1 vacancy per 3200 Pb atoms assuming no overlap. More details about the irradiation experiments can be seen in our previous publications \cite{Taenirra,SunPhysRevB.96.140505irra,ParkPhysRevB.98.054512,Sun_APEXHirraFeSe}.

PbTaSe$_2$ consists of stacking of hexagonal TaSe$_2$ layers intercalated with Pb. Fig. 1(a) shows the crystal structure looking down the hexagonal TaSe$_2$ plane. The intercalated Pb atoms sit above the Se atoms, and make the crystal structure noncentrosymmetric as shown in the side view [Fig. 1(b)] \cite{AliPRB,BianPbTaSeNatComm}. $\phi$ defines the azimuthal angle of the magnetic field rotates in the hexagonal plane [Fig. 1(a)], while $\theta$ defines the polar angle of the magnetic field away from the $c$-axis [Fig. 1(b)].

Fig. 1(c) shows the temperature dependence of magnetic susceptibility, which displays $T_{\rm{c}}$ $\sim$ 3.7 K ($T_{\rm{c}}$ is defined by the onset of the deviation between the zero-field-cooling and field-cooling susceptibilities.) with a sharp transition width less than 0.2 K by taking the criteria of 10\% and 90\% of the magnetization result at 1.8 K. The $T_{\rm{c}}$ is also confirmed by the zero resistivity [see the inset of Fig. 1(d)]. The main panel of Fig. 1(d) shows the temperature dependence of the in-plane resistivity at the temperature range from 320 K to 2 K. The residual resistivity ratio RRR defined as $\rho$(300 K)/$\rho$($T_{\rm{c}}^{onset}$) is estimated as large as $\sim$ 321. The sharp SC transition width and the large RRR confirm the high-quality of the single crystal. 

Fig. 1(e) shows the temperature dependence of specific heat divided by temperature $C/T$ under various magnetic fields ($H\parallel c$ ) ranging from 0 to 1000 Oe. A clear jump associated with superconducting transition is observed at around 3.7 K under zero field, which is consistent with the susceptibility and resistivity measurements. The SC jump is gradually suppressed by magnetic field, and no SC jump can be observed down to $\sim$ 0.4 K under $H$ = 800 Oe. On the other hand, an upturn can be observed at temperatures below $\sim$ 0.7 K when $H$ $>$ 600 Oe. The NMR measurements have proved that electron-electron interaction is very weak in this material based on the observation that the spin-lattice relaxation rate varies in proportion to the temperature \cite{MaedPbTaSePRB}. Therefore, such upturn behavior may be from the Schottky behavior of nuclear contribution. 

The normal state specific heat can be fitted by the sum of electronic part and phononic part: $C_n/T$ = $\gamma_n$ + $\beta_nT^2$ + $\alpha_nT^4$. The fitting result is shown as the solid line in Fig. 1(e) giving $\gamma_n$ = 8.56 mJ/mol$\cdot$K$^2$, $\beta_n$ = 2.12 mJ/mol$\cdot$K$^4$, and $\alpha_n$ = 0.028 mJ/mol$\cdot$K$^6$. The validity of the fitting is justified by the entropy conservation as shown in the inset of Fig. 1(e). Besides, a very small residual specific heat $\gamma_0$ under zero-field at 0 K is obtained as $\sim$ 0.5 mJ/mol$\cdot$K$^2$ by linearly extrapolating the data of $C/T$ vs $T^2$ down to 0 K [see supplementary Fig. S1 \cite{supplement}]. The normalized specific heat jump at $T_{\rm{c}}$, $\Delta C/(\gamma_n-\gamma_0)T_c$ is estimated to be 1.39, which is close to the weak-coupling value 1.43 of BCS theory. A similar value of the normalized specific heat jump is also reported in the previous report \cite{AliPRB}, while a slightly larger value $\sim$ 1.71 is also reported \cite{ZhangPbTaSePRB}. Zero-field electronic specific heat $C_{\rm{e}}/T$ obtained from subtracting the phonon terms, is shown in Fig. 1(f). $C_{\rm{e}}/T$ decreases to a small value close to zero at low $T$, indicating the nodeless gap structure similar to previous reports \cite{ZhangPbTaSePRB,WangPbTaSePRB}. On the other hand, $C_{\rm{e}}/T$ increases linearly with $T$ in a wide temperature region above $T$ = 1 K, which is different from the exponential increase for an single-gapped isotropic $s$-wave superconductor. The fitting of $C_{\rm{e}}$ will be discussed later.

\begin{figure*}\center
	\includegraphics[width=14cm]{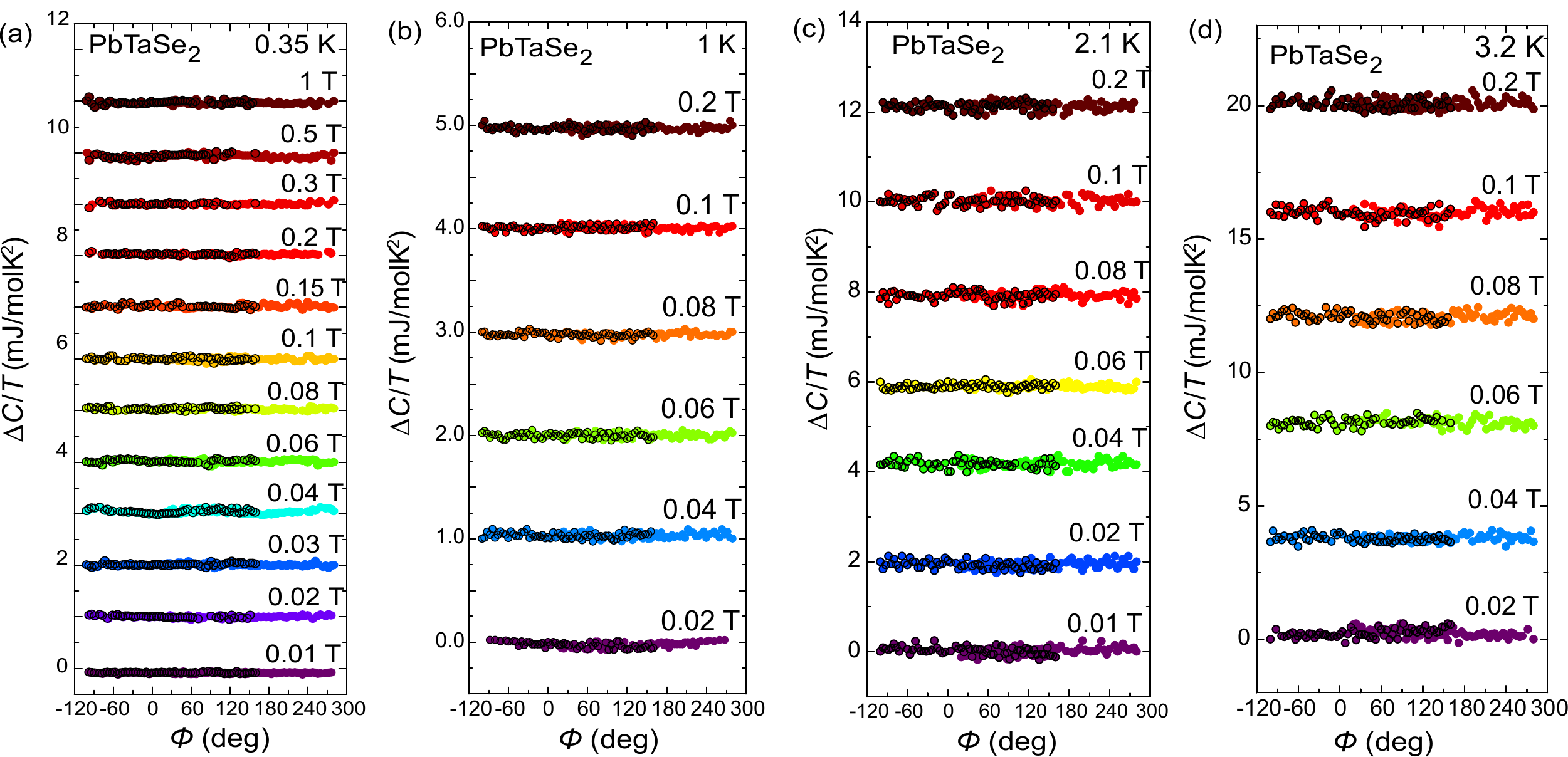}\\
	\caption{Azimuthal angle dependence of the specific heat $\Delta C(\phi)/T$ measured under various fields at (a) 0.35 K, (b) 1 K, (c) 2.1 K, and (d) 3.2 K, respectively. Each subsequent curve at the same temperature is shifted vertically to show the angle dependence more clearly. Black-outlined symbols are the measured data; the others are mirrored points to show the angle dependence more clearly.}\label{}
\end{figure*}

\begin{figure*}\center
	\includegraphics[width=14cm]{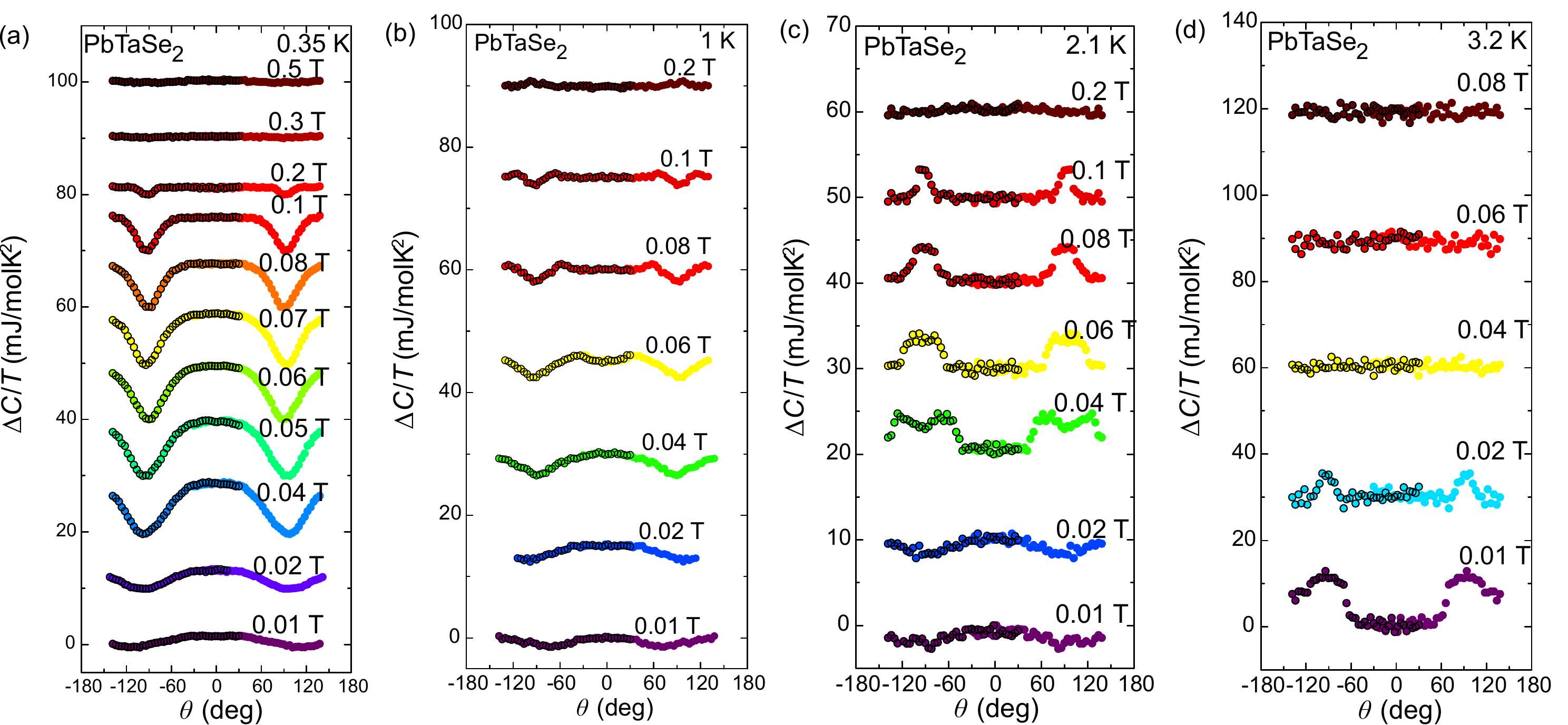}\\
	\caption{Polar angle dependence of the specific heat $\Delta C(\theta)/T$ measured under various fields at (a) 0.35 K, (b) 1 K, (c) 2.1 K, and (d) 3.2 K, respectively. Each subsequent curve at the same temperature is shifted vertically to show the angle dependence more clearly. Black-outlined symbols are the measured data; the others are mirrored points to show the angle dependence more clearly.}\label{}
\end{figure*}

More information about the gap structure can be obtained from the magnetic field dependence of the specific heat, $C/T$ vs $H$, which usually reflects the QPs excitation within the SC gap. For a superconductor with an isotropic single gap, $C$($H$)/$T$ is linearly related to $H$ ($C$($H$) $\sim$ $C_{\rm{n}}$($H$/0.8$H_{c2}$) under small field, where $C_{\rm{n}}$ is the normal state specific heat, $H_{c2}$ is the upper critical field) because the low-energy QPs are mainly localized in the vortex core, whose density is proportional to $H$ \cite{NakaiPhysRevB.70.100503}. For a single gap with nodes, $C$($H$)/$T$ is not linearly proportional to $H$ ($\propto H^{1/2}$ for the line nodes, and $\propto H^{0.64}$ for the point nodes \cite{MiranovicPhysRevB.68.052501}) because of the QPs' Doppler shift caused by supercurrents around the vortex core \cite{VolovikJETPLett}. In between, $C$($H$)/$T$ for an anisotropic single gap or multi-gaps shows a crossover from linear-$H$ to $H^{1/2}$ dependence \cite{NakaiPhysRevB.70.100503}. The $C$($H$)/$T$ of PbTaSe$_2$ at 0.42 K for $H \parallel c$ is shown in Fig. 1(g), which resides between the linear and $H^{1/2}$ behaviors. This result excludes the single isotropic gap structure, the line or loop nodes (both vertical or horizontal), and the point nodes in the $ab$-plane. However, the point nodes along $c$ axis cannot be excluded by the measurements with $H \parallel c$ because the low-energy QPs will not be excited when the field is parallel to the nodal direction. Besides, the anisotropic single gap and multi-gaps cannot be distinguished in the $C$($H$)/$T$ results. 

Therefore, only with the temperature and field dependent specific heat results, we cannot reveal the angle-resolved gap structure of PbTaSe$_2$. According to the Doppler shift effect, $\delta E = m_e \textbf{\emph{v}}_F \cdot \textbf{\emph{v}}_s$ ($m_e$ is the electron mass, $\emph{\textbf{v}}_F$ is the Fermi velocity, and $\emph{\textbf{v}}_s$ is the local superfluid velocity always perpendicular to the field) \cite{VolovikJETPLett}, the zero-energy DOS under small fields in superconductors with nodes or gap minima depends on the direction of the field with respect to the nodal/gap minimum positions. When $H\parallel$ nodes/gap minima, it shows minima because $\delta E=0$ in the case of $\emph{\textbf{v}}_F\perp\emph{\textbf{v}}_s$, while it turns into maxima when $H\perp$ nodes/gap minima because $\delta E$ becomes maximal in the situation of $\emph{\textbf{v}}_F\parallel\emph{\textbf{v}}_s$. Therefore, in the low-field region, specific heat shows minima for $H\parallel$ node/gap minimum, and maxima for $H\perp$ node/gap minimum. On the other hand, under high fields, the QP scattering by the magnetic field is strongly enhanced, exciting much higher finite-energy DOS around the nodal positions for $H\parallel$ node/gap minimum. When the finite-energy DOS overcomes the zero-energy DOS, the oscillation switches signs, i.e., specific heat becomes maxima for $H\parallel$ node/gap minimum, but minima for $H\perp$ node/gap minimum \cite{VorontsovPhysRevLett.96.237001,HiragiJPSJASHcal}. Such sign change behavior has been observed in superconductors with nodes \cite{AnPhysRevLett.104.037002,KittakaKFe2As2JPSJ} or gap minima \cite{YueSunPRBFeSeARSH,SunPhysRevB.98.064505}. By contrast, for the isotropic gap structure, the specific heat should be independent of the field direction. To reveal the angle-resolved gap structure of PbTaSe$_2$, we turn to the measurements of specific heat under magnetic field with angle resolution.

Figures 2(a)-2(d) show the azimuthal angle-resolved specific heat $\Delta C(\phi)/T$ under different magnetic fields at 0.35 K, 1 K, 2.1 K, and 3.2 K, respectively. Obviously, no oscillation is observed at any temperature and field. Usually the magnitude of the oscillation from nodes/gap minima is about a few percent of the total electronic specific heat, $C_{\rm{e}}(H)$ \cite{SakakibaraReview,YueSunPRBFeSeARSH}. In supplementary Fig. S2, we show that the noise level of our system is $\sim$ 0.07 mJ/mol$\cdot$K$^2$ at 0.35 K and 0.08 T. In this case, an oscillatory signal (e.g 2-fold anisotropy) in the $\Delta C(\phi)/T$ with an amplitude of 1$\sim$3\% of $C_{\rm{e}}/T$ should still be observed if it exists [see supplementary material S2 \cite{supplement}]. Therefore, our azimuthal angle-resolved specific heat results confirm that the in-plane gap structure should be nearly isotropic, although we cannot exclude the possibility of very tiny anisotropy below our resolution limit. On the other hand, the out-of-plane gap structure cannot been resolved only by the azimuthal angle-resolved measurements, because the point nodes/gap minima along $c$ axis will not cause oscillation in $\Delta C(\phi)/T$ \cite{TsutsumiPhysRevB.94.224503}.    

In order to probe the gap structure along $c$ axis, and reveal the 3D gap structure of PbTaSe$_2$, we also measured the polar-angle dependence of the specific heat $\Delta C(\theta)/T$ [see Fig. 3]. If the point nodes/gap minima along $c$-axis are present, the $\Delta C(\theta)/T$ is expected to show an oscillatory behavior with sign change as discussed above. Nevertheless, $\Delta C(\theta)/T$ does not display such features. Instead, it manifests a twofold symmetry at low temperature such as 0.35 K under small fields [see Fig. 3(a)], which is simply attributed to the out-of-plane anisotropy of $H_{c2}$. Under larger fields, $\Delta C(\theta)/T$ becomes flat ($\theta$-independent) at the $\theta$-range around 0, which comes from the suppression of SC for $H\parallel c$  ($\theta$ = 0) due to the small $H_{c2}$. Subsequently, the flattened region evolves into wider $\theta$-range with further increasing field. Finally the $\Delta C(\theta)/T$ becomes totally $\theta$-independent at 0.3 T, indicating that the $H_{c2}$ is reached also for $H\parallel ab$ ($\theta$ = $\pm$90$^\circ$). The out-of-plane anisotropy is consistent to that obtained from transport and magnetic susceptibility measurements \cite{ZhangPbTaSePRB,SankarPbTaSeJPCM}.

On the other hand, at higher temperatures, $\Delta C(\theta)/T$ shows some special features such as the peaks when $H\parallel ab$ ($\theta$ = $\pm$90$^\circ$) at 2.1 K for $H\geq$ 0.04 T [see Fig. 3(c)]. These features are different from the sign change behavior induced by nodal gap structure. In the $\theta$-range around 0$^\circ$, $\Delta C(\theta)/T$ at 2.1 K already becomes $\theta$-independent for $H\geq$ 0.04 T, which suggests that it is in the normal state. Meanwhile, $\Delta C(\theta)/T$ around $\pm$90$^\circ$ stays in the SC state. The larger $\Delta C(\theta = \pm90^\circ)/T$ than $\Delta C(\theta = 0^\circ)/T$ originates from the larger specific heat in the SC state close to $T_{\rm{c}}$ than that in the normal state. The dips at $\theta = \pm90^\circ$ observed at 2.1 K and 0.04 T is due to the crossover from minimum to peak.

Here, we emphasize that the relatively large out-of-plane anisotropy induces a strong oscillation in $\Delta C(\theta)/T$, which may cover the feature from possible $c$-axis nodes such as the sign change behavior. Therefore, we cannot simply exclude the $c$-axis point nodes only by the $\Delta C(\theta)/T$ results. To have more evidence, we also measured the $C$($H$)/$T$ with $H \parallel ab$ at 0.33 K as shown in the Fig. S3 \cite{supplement}. The field dependence of $C$($H$)/$T$ obviously deviates from the expected behavior of $\propto H^{0.64}$ for the point nodes \cite{MiranovicPhysRevB.68.052501}. It increases linearly with magnetic field with different slopes in the low field and high field regions, representing the behavior of two-gap superconductor. Those results confirm that there is no $c$-axis point nodes in PbTaSe$_2$. 

Now, we return to the temperature dependence of $C_{\rm{e}}/T$ shown in Fig. 1(f). Based on the above discussions, we fit the $C_{\rm{e}}/T$ with a two-gap model based on the BCS theory by simply assuming $\Delta_1$ and $\Delta_2$ are both isotropic $s$-waves. In this case, $C_{\rm{e}}$=$\eta_1 C_2(\Delta_1)$+$\eta_2 C_2(\Delta_2)$, where $C_i$ denotes the electronic specific heat for each gap, while $\eta_i$ denotes the ratio of each gap. The data are well-fitted, as shown by the solid line in Fig. 1(f), with the gap values $\Delta_1$ = 0.28 meV and $\Delta_2$ = 0.58 meV, and the relative weight of $\eta_1$ = 16\%, and $\eta_2$ = 84\%. A two-gap model with two isotropic $s$-wave gap has been also applied in the fitting of conductance curve of STM \cite{GuanPbTaSeSciAdv}.  

\begin{figure}\center
	\includegraphics[width=8.5cm]{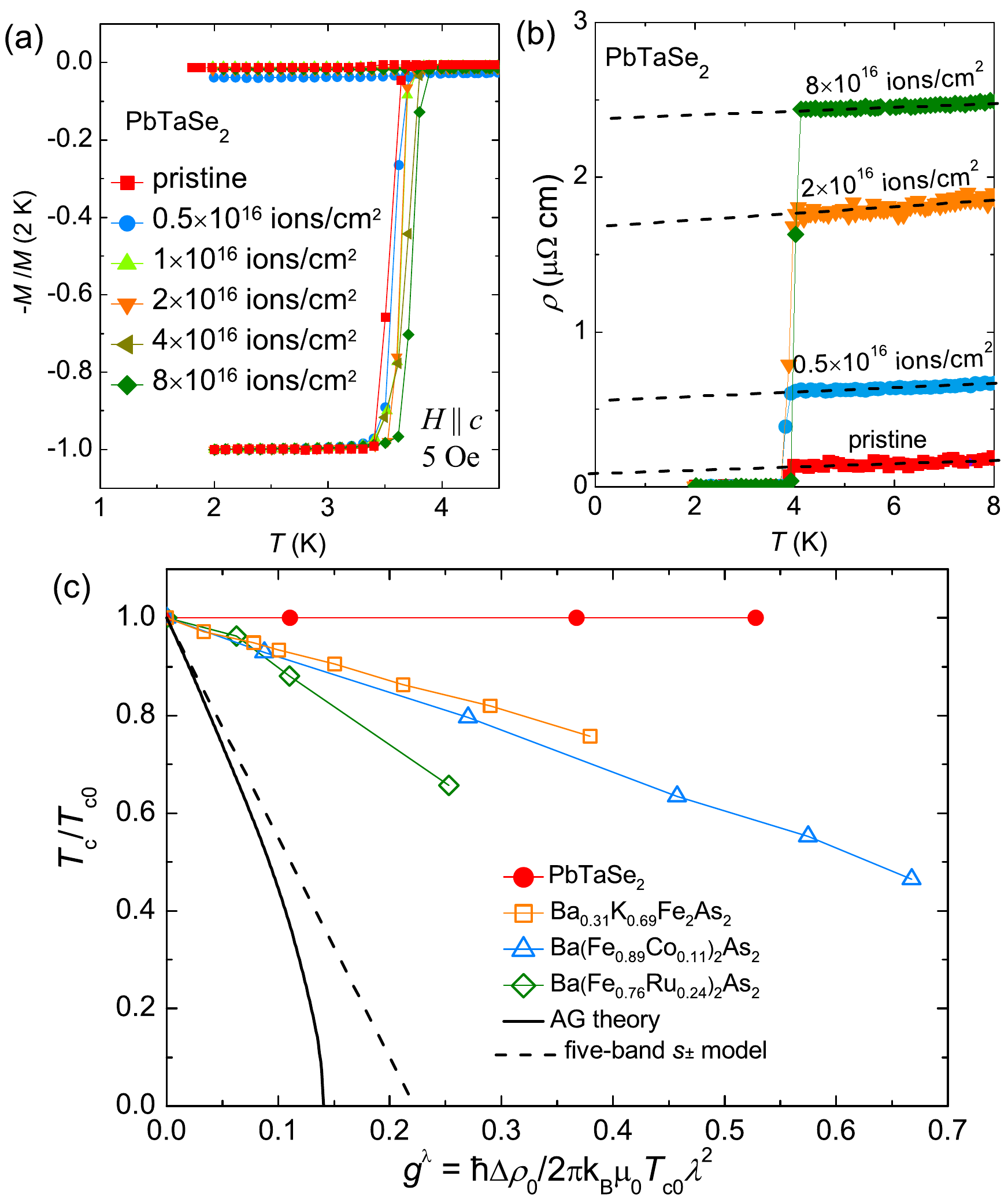}\\
	\caption{(a) Temperature dependence of the normalized magnetization at 5 Oe for the pristine and H$^+$-irradiated crystals. (b) Temperature dependence of the resistivity for the pristine and some selected irradiated crystals. Dashed lines are the linear extrapolations to zero temperature for estimating the residual resistivity $\rho_0$. (c) $T_{\rm{c}}$/$T_{\rm{c0}}$ as a function of a dimensionless scattering rate $g^{\lambda}$ = $\hbar\Delta\rho_0$/2$\pi k_B\mu_0T_{c0}\lambda^2$. The solid line is the $T_{\rm{c}}$ suppression rate predicted by the Abrikosov-Gor’kov (AG) theory for an isotropic $s$-wave superconductor with magnetic impurities \cite{AGtheory}. The dashed line represents the theoretical prediction by a five-band $s_{\pm}$ model \cite{OnariPRLdisorder}. For comparison, we also plot the results from the Ba$_{0.31}$K$_{0.69}$Fe$_2$As$_2$ \cite{Taenirra}, Ba(Fe$_{0.89}$Co$_{0.11}$)$_2$As$_2$ \cite{NakajimaPhysRevB.82.220504}, and Ba(Fe$_{0.76}$Ru$_{0.24}$)$_2$As$_2$ \cite{ProzorovPhysRevX.4.041032} with possible $s_{\pm}$ pairing.}\label{}
\end{figure}

Until now, the temperature, field, and angle dependences of specific heat have confirmed that the bulk gap structure of PbTaSe$_2$ consists of two 3D isotropic $s$-wave. However, even for the $s$-wave gap structure, the gap function can be sign reversed or preserved. The sign-reversed $s$-wave, usually called as the $s_{\pm}$, has been suggested to be the most plausible gap function of iron-based superconductors \cite{Mazinnature,HirschfeldReportsonProgressinPhysics}. The amount of excited QPs should be the same for both the sign reversed and preserved $s$-wave. Therefore, to distinguish the two possible gap functions, a phase-sensitive technique is needed. The nonmagnetic disorder effect induced by light-particle irradiation, such as the electron and H$^+$, has been proven to be an effective method to discriminate the sign-reversed $s$-wave from the and sign-preserved one. For the sign-reversed $s_{\pm}$, $T_{\rm{c}}$ is usually suppressed remarkably by the nonmagnetic disorders, while $T_{\rm{c}}$ for the sign-preserved $s$-wave is robust \cite{WangPRBdisorder,OnariPRLdisorder}.       

The H$^+$-irradiation effect on PbTaSe$_2$ is summarized and presented in Fig. 4. It should be noted that PbTaSe$_2$ contains the topological surface state, whose gap function is reported to be different from the bulk one, which may have nodes \cite{LeTianarXiv,Hui-KeJinarXiv}. To avoid the possible influence of surface state, $T_{\rm{c}}$ for the H$^+$-irradiated crystals were probed by both the magnetic susceptibility and resistivity measurements as shown in Figs. 4(a) and 4(b), respectively. Obviously, the irradiated crystals display nearly the same $T_{\rm{c}}$ to the pristine one, which proves that there is no node in the gap structure, otherwise $T_{\rm{c}}$ will be suppressed with smearing the nodes by introducing disorders. This result is consistent to the observation from specific heat. 

For quantitative discussion of the pairing-breaking effect and comparison with other superconductors, the $T_{\rm{c}}$ suppression rate $T_{\rm{c}}$/$T_{\rm{c0}}$ is plotted against the dimensionless scattering rate $g^{\lambda}$ as shown in Fig. 4(c), where $T_{\rm{c}}$ is obtained from the resistivity measurements shown in Fig. 4(b), and $T_{\rm{c0}}$ is the value of $T_{\rm{c}}$ for the pristine one. $g^{\lambda}$ = $\hbar\Delta\rho_0$/2$\pi k_B\mu_0T_{c0}\lambda^2$ \cite{KoganPhysRevB.80.214532}, where $\hbar$ is the Planck’s constant divided by 2$\pi$, $k_B$ is the Boltzmann constant, $\mu_0$ is the vacuum permeability, and $\lambda$ is the penetration depth $\backsimeq$ 204 nm evaluated from a tunnel diode oscillator experiment \cite{PangPRBpenetration}. $\Delta\rho_0$ is the increase of residual resistivity after irradiation, $\Delta\rho_0$ = $\rho_0^{\rm{irradiated}}$ - $\rho_0^{\rm{pristine}}$. The residual resistivity $\rho_0$ was obtained by linearly extrapolating $\rho$-$T$ curves in the normal state above $T_{\rm{c}}$ to $T$ = 0 K, as shown by the dashed lines in Fig. 4(b). In the plot of $T_{\rm{c}}$/$T_{\rm{c0}}$ vs $g^{\lambda}$, the disorder level is expressed by the $\rho_0$, evaluated simultaneously with $T_{\rm{c}}$ in the $\rho$-$T$ measurements, which avoids the influence from possible thermal annealing after irradiation. Indeed, the increase of $\rho_0$ for the crystal irradiated by 8$\times$10$^{16}$/cm$^2$ of H$^+$ is smaller than expected, indicating the possible annealing effect before measurements after irradiation.                            

For comparison, we also plot the $T_{\rm{c}}$ suppression rates expected by the Abrikosov-Gor’kov (AG) theory for an isotropic $s$-wave superconductor with magnetic impurities (solid line) \cite{AGtheory}, and that by a five-band $s_{\pm}$ model (dashed line) \cite{OnariPRLdisorder}, together with those from Ba$_{0.31}$K$_{0.69}$Fe$_2$As$_2$ \cite{Taenirra}, Ba(Fe$_{0.89}$Co$_{0.11}$)$_2$As$_2$ \cite{NakajimaPhysRevB.82.220504}, and Ba(Fe$_{0.76}$Ru$_{0.24}$)$_2$As$_2$ \cite{ProzorovPhysRevX.4.041032} with possible $s_{\pm}$ pairing. Clearly, the $T_{\rm{c}}$ suppression rate for PbTaSe$_2$ is much smaller than those theories and materials. Recently, the sign-preserved pairing is discussed on the heavy-fermion CeCu$_2$Si$_2$ based on the observation of robust superconductivity against impurities \cite{YamashitScienceAdvances}. On the other hand, similar behavior of the unchanged $T_{\rm{c}}$ against H$^+$- and electron-irradiations has been observed in an established two-gap $s_{++}$ superconductor MgB$_2$ \cite{KleinPhysRevLett.105.047001,MezzettiprotonMgB2}. Therefore, such robust $T_{\rm{c}}$ to the disorders demonstrates that the gap function of PbTaSe$_2$ should be sign-preserved rather than sign-reversed. Thus, our combined ARSH and disorder effect studies have proved that PbTaSe$_2$ consists two isotropic $s$-wave gaps without sign-reversal. 
  
In summary, we have investigated the bulk superconducting gap function of topological superconductor PbTaSe$_2$ through the combined studies of ARSH and disorder effect. Temperature, magnetic field, and 3D angle dependences of specific heat measurements prove that the superconductivity of PbTaSe$_2$ is fully-gapped, with two isotropic $s$-wave gaps. The superconducting transition temperature $T_{\rm{c}}$ is found to be robust against non-magnetic disorders induced by H$^+$-irradiation, which suggests that the gap functions should have the same sign.

This work was partly supported by JSPS KAKENHI (Nos. JP20H05164, JP19K14661, 18H05853, 18H01161, 18H04306, 17K05553, 17H01141, and 15H05883), and the National Natural Science Foundation of China (Grant Nos. U1932217, 11674054, U1732162). RS acknowledges financial support provided by the Ministry of Science and Technology in Taiwan under (No. MOST-108-2112-M-001-049-MY2).


\bibliography{PbTaSereferences}

\pagebreak

\newpage
\onecolumngrid
\begin{center}
	\textbf{\huge Supplemental information}
\end{center}
\vspace{1cm}
\twocolumngrid
\setcounter{equation}{0}
\setcounter{figure}{0}
\setcounter{table}{0}

\makeatletter
\renewcommand{\theequation}{S\arabic{equation}}
\renewcommand{\thefigure}{S\arabic{figure}}

\subsection*{S1 Estimation of the residual specific heat}

\begin{figure}\center
	\includegraphics[width=8.5cm]{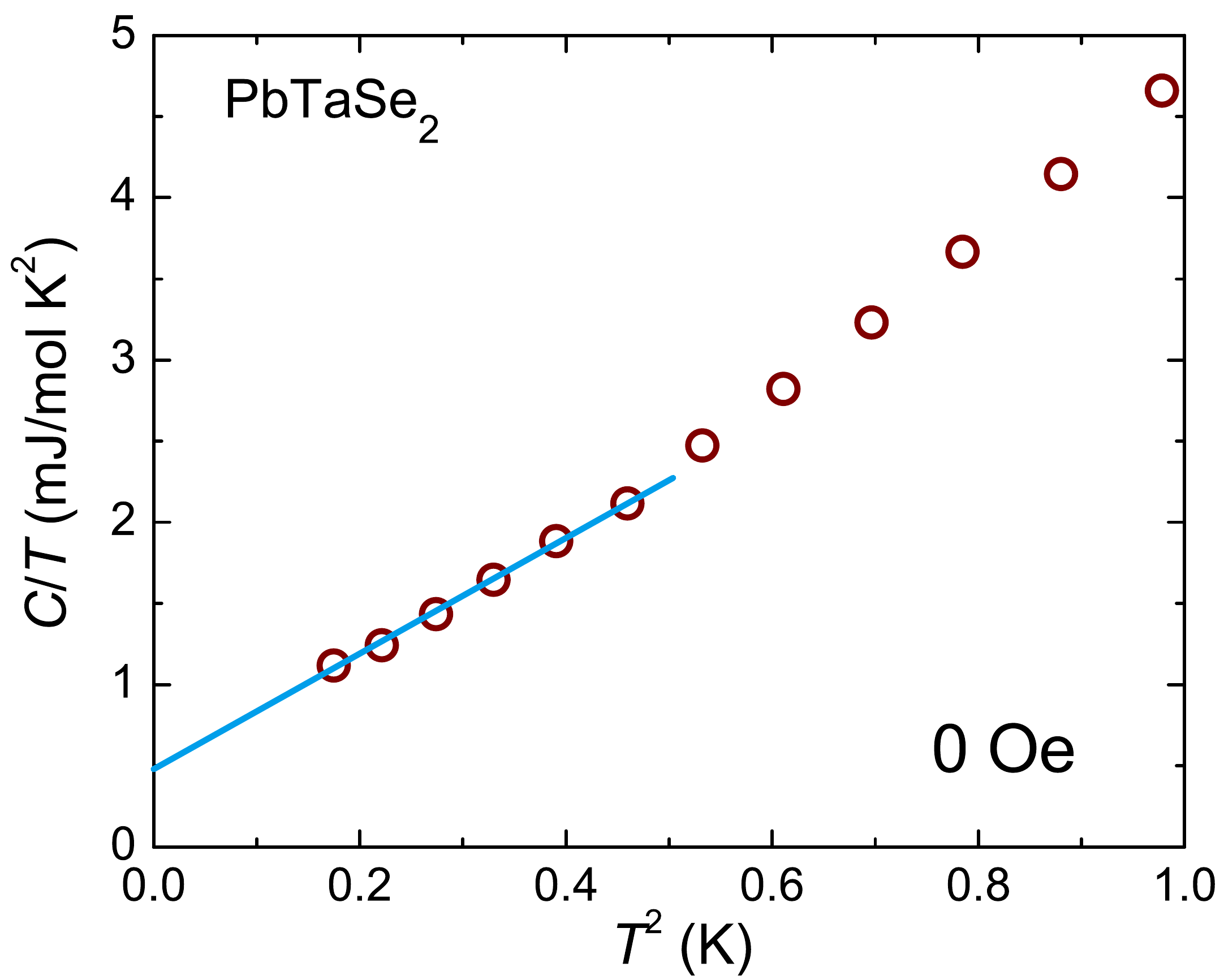}\\
	\caption{Temperature dependence of specific heat under zero field plotted as $C/T$ vs $T^2$ below 1 K. The solid line represents the linear extrapolation of the data down to 0 K.}\label{}
\end{figure}

Figure S1 shows the specific heat divided by temperature $C$/$T$ as a function of $T^2$ under zero magnetic field. The solid line represents the linear extrapolation of the data down to 0 K. The residual specific heat $\gamma_0$ is estimated as $\sim$ 0.5 mJ/mol$\cdot$K$^2$.

\subsection*{S2 Noise level of the azimuthal angle dependence of specific heat}

\begin{figure}\center
	\includegraphics[width=8.5cm]{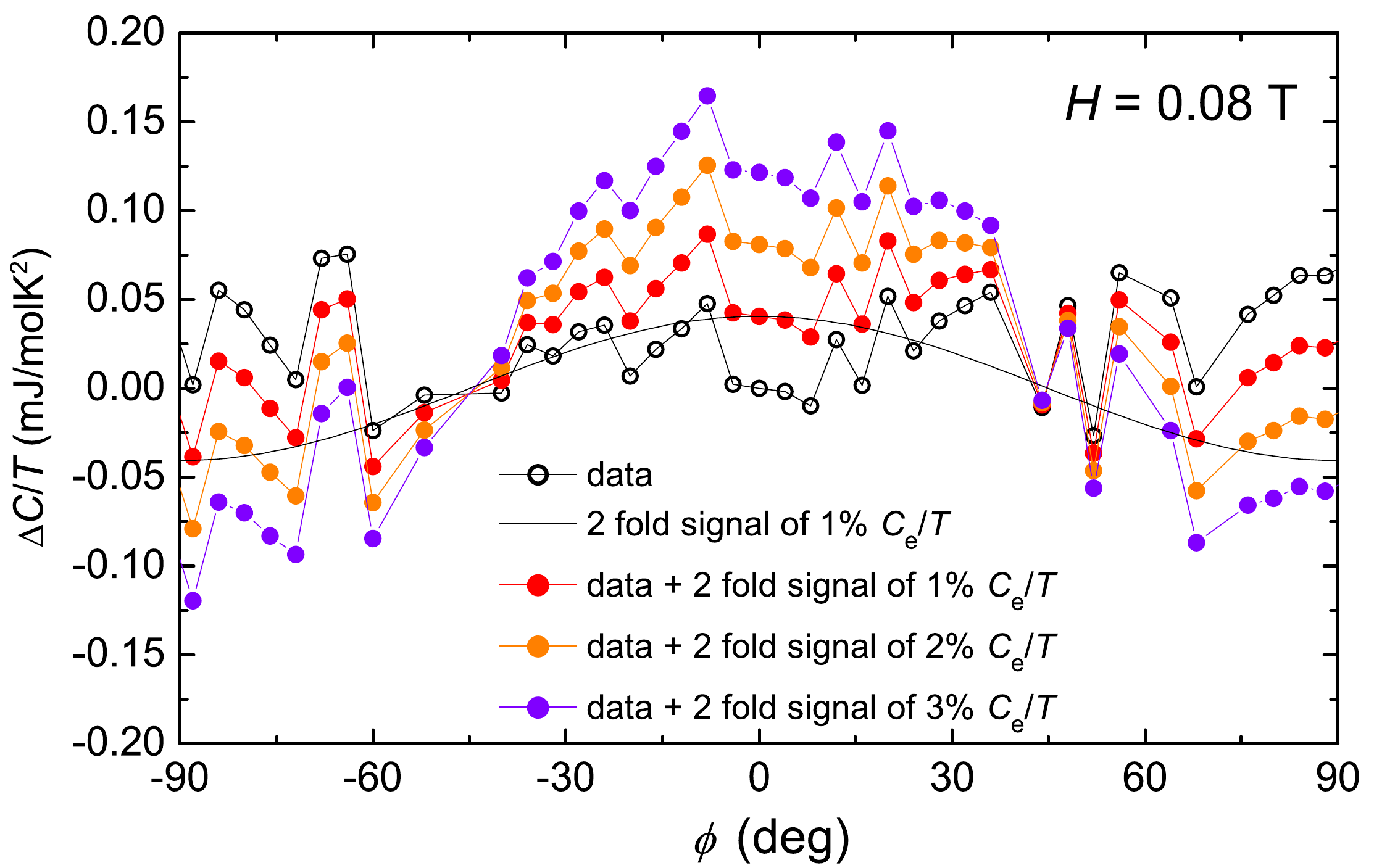}\\
	\caption{Azimuthal angle dependence of specific heat $\Delta C$/$T$ under 0.08 T at 0.35 K, the 2-fold signal of 1\% of $C_{\rm{e}}/T$ at 0.08 T, and the superposition of the data with 1\%, 2\%, and 3\% of the 2-fold signal of $C$/$T$.}\label{}
\end{figure}

Figure S2 shows the azimuthal angle dependence of specific heat under 0.08 T at 0.35 K (open black circles) together with the putative 2-fold signal of 1\% of $C_{\rm{e}}/T$ (solid line). It is obvious that the noise level is around 0.07 mJ/mol$\cdot$K$^2$. In this case, the fluctuation of the signal with respect to ($\gamma_n$-$\gamma_0$) $\sim$ 8 mJ/mol$\cdot$K$^2$ is $\sim$ 0.07/8 = 0.88\%. Superpositions of the data with putative 2-fold signals with amplitudes of 1\%, 2\%, and 3\% of $C_{\rm{e}}/T$ are also shown in Fig. S2 in red, orange, and purple solid circles, respectively. Even though the amplitude of 2-fold signal is comparable to the fluctuation of the data, it can be observed if it really exists.  

\subsection*{S3 Magnetic field dependence of specific heat at 0.33 K for $H \parallel ab$}

\begin{figure*}\center
	\includegraphics[width=14cm]{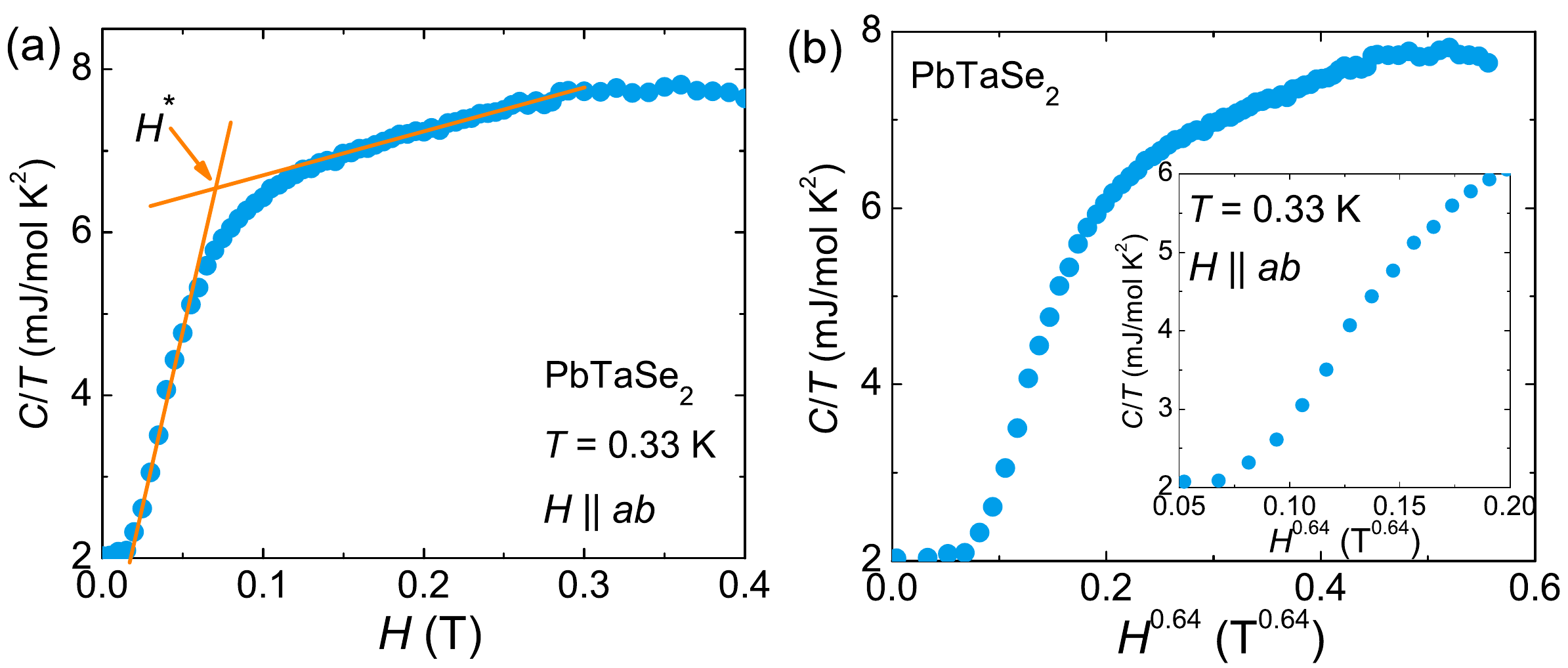}\\
	\caption{(a) Magnetic field dependence of specific heat at 0.33 K for $H \parallel ab$-plane. (b) The plot of $C$/$T$ vs $H^{0.64}$, and the inset is the enlarged range of 0.05 T$^{0.64}$ $< H^{0.64} <$ 0.2 T$^{0.64}$.}\label{}
\end{figure*}

To check the possible $c$-axis point nodes, we performed the measurements of field dependent specific heat with $H \parallel ab$-plane. (The measurement was done on another piece of crystal.) $C$/$T$ increases linearly with magnetic field with different slopes in the low field and high field regions [see Fig. S3(a)], representing the behavior of two-gap superconductor. It is clearly different from the expected behavior of $C$/$T$ $\propto H^{0.64}$ for point nodes as shown in Fig. S3(b), and the enlarged plot in the range of 0.05 T$^{0.64}$ $< H^{0.64} <$ 0.2 T$^{0.64}$ [see inset of Fig. S3(b)]. Therefore, the $c$-axis point nodes can be excluded. 

For a typical two-gap superconductor, the linear increase of $C$/$T$ in the low field region is dominant by the suppression of the gap with smaller upper critical field, which is usually defined as the virtual upper critical field $H^{\ast}$ as shown in Fig. S3(a). After the magnetic field increased above $H^{\ast}$, $C$/$T$ will linearly increase with field at the high field region due to the suppression of another gap with large upper critical field. The virtual upper critical field $H^{\ast}$ of PbTaSe$_2$ is estimated as 0.07 T as shown in Fig. S3(a). The value of $C$/$T$ at $H^{\ast}$ is $\sim$ 6.5 mJ/mol$\cdot$K$^2$, which is about 83\% of the $\gamma_n$ $\sim$ 7.8 mJ/mol$\cdot$K$^2$, which is consistent to the ratio of the two gaps (84\% : 16\%) obtained by the fitting of $C_e/T$ [see Fig. 1(f)]. According to the $C_e/T$ fitting, the larger ratio (84\%) corresponds to the larger gap $\Delta_2$. Therefore, $H^{\ast}$ is the upper critical field for $\Delta_2$.

On the other hand, the virtual upper critical field $H^{\ast}$ can be expressed as $H^{\ast}$ $\sim$ $\Phi_0/2\pi\xi_{ab}^{\ast}\xi_c^{\ast}$, while the upper critical filed for $H \parallel ab$ can be expressed as $H_{c2}^{ab}$ $\sim$ $\Phi_0/2\pi\xi_{ab}\xi_c$. Here, $\xi_{ab}^{\ast}$ = $\hbar v_{F2}^{ab}$/$\pi\Delta_2$, and $\xi_{c}^{\ast}$ = $\hbar v_{F2}^{c}$/$\pi\Delta_2$ are the coherence length along $ab$ plane and $c$ axis for the larger gap $\Delta_2$ (0.58 meV). $\xi_{ab}$ = $\hbar v_{F1}^{ab}$/$\pi\Delta_1$, and $\xi_{c}$ = $\hbar v_{F1}^{c}$/$\pi\Delta_1$ are the coherence length along $ab$ plane and $c$ axis for the smaller gap $\Delta_1$ (0.28 meV). $v_{Fi}$ ($i$ = 1, 2) is the Fermi velocity for each band. Then, the ratio of $H^{\ast}$/$H_{c2}^{ab}$ can be expressed as 
\begin{equation}
\label{eq.1}
\frac{H^{\ast}}{H_{c2}^{ab}}=\frac{\xi_{ab}\xi_c}{\xi_{ab}^{\ast}\xi_c^{\ast}}=\frac{\Delta_2^2}{\Delta_1^2}\frac{v_{F1}^{ab}v_{F1}^{c}}{v_{F2}^{ab}v_{F2}^{c}}.
\end{equation}                    
The $H_{c2}^{ab}$ is obtained as $\sim$ 0.3 T from Fig. S3(a). The ratio of Fermi velocity for the two bands can be estimated as $v_{F2}$/$v_{F1}$ = 4.3, assuming the isotropic Fermi velocity $v_{Fi}^{ab}$ = $v_{Fi}^c$. Such large difference in the Fermi velocity between different bands may originate from the topological band structure of PbTaSe$_2$, where the large $v_{F2}$ is from the Dirac band with linear dispersion, while the small $v_{F1}$ is from the traditional parabolic band. Actually, a large difference in the effective mass in different bands has been observed by the quantum oscillation measurements [20].

\end{document}